# Dataset Quality Assessment: An extension for analogy based effort estimation

Mohammad Azzeh

Department of Software Engineering, Applied Science University, PO BOX 166, Jordan

m.y.azzeh@asu.edu.jo

**Abstract**
Estimation by Analogy (EBA) is an increasingly active research method in the area of software engineering. The fundamental assumption of this method is that the similar projects in terms of attribute values will also be similar in terms of effort values. It is well recognized that the quality of software datasets has a considerable impact on the reliability and accuracy of such method. Therefore, if the software dataset does not satisfy the aforementioned assumption then it is not rather useful for EBA method. This paper presents a new method based on Kendall's row-wise rank correlation that enables data quality evaluation and providing a data pre-processing stage for EBA. The proposed method provides sound statistical basis and justification for the process of data quality evaluation. Unlike Analogy-X, our method has the ability to deal with categorical attributes individually without the need for partitioning the dataset. Experimental results showed that the proposed method could form a useful extension for EBA as it enables: dataset quality evaluation, attribute selection and identifying abnormal observations.

## 1. Introduction

Software development projects are highly complex processes involving various kinds of risks **[13, 18, 19]**. Thus, delivering a high quality software product on time and within budget requires a reliable and accurate software effort estimation method **[10]**. EBA is a commonly used method and viable alternative to other conventional estimation methods **[3, 16]**. EBA is an analogical reasoning technique that aims at identifying within a historical case base the source projects that are similar to the target project in terms of project description (i.e. attributes) for which their solutions are reused to generate a new solution **[13, 24]**.

The quality of historical dataset is the key component of any effort estimation method **[17]**. The quality in the context of EBA method means that the dataset should significantly satisfy the fundamental EBA assumption which is "*the projects that are similar in terms of their attributes values will also be similar in terms of their effort values*". **[24]** The aforementioned assumption implies that any attribute is regarded as useful and reliable for prediction only if the ranks of similar source projects with respect to that attribute values should be consistent with ranks of similar source projects with respect to their effort values. This assumption can be tested using Kendall's row-wise rank correlation between similarity matrix based project attribute values and similarity matrix based effort values.

ANGEL **[24]**, which is the most well known EBA system, offers a flexible tool and support for various kinds of attribute selection algorithms and validation techniques within a user friendly GUI tool. Even though, it has no method to test the appropriateness of datasets for EBA. The current method for identifying reliable attribute uses brute-force search which is almost optimized based upon some performance indicators such as MMRE or PRED. These performance indicators are widely regarded as problematic and have many inadequacies as discussed in **[11]**. In addition, the selected attributes depend on the desired configuration of analogy-based system such as similarity measure, adaptation rules and number of analogies.

Recently, Keung et al. **[16]** proposed a method called Analogy-X to statistically test the hypothesis of EBA using Mantel correlation. The method utilizes Mantel correlation between distance matrix of attribute values and distance matrix of effort values, following the stepwise procedure. This method has a certain limitation as stated and confirmed by Keung et al. **[16]**, in that it cannot directly deal with the distance matrix of nominal data individually. So it cannot assess the appropriateness of nominal attribute solely, but alternatively by dividing the best continuous attributes into subgroups according to the number of categories in that nominal attribute. This solution removes large number of elements in the matrix, because only matrix elements within the same category are considered in the correlation calculation. As result, the use of nominal attribute requires special treatment, largely due to the nature of similarity matrices formed by categorical variables **[16]**.

In the present paper, we propose a new method that enables us to assess the quality and reliability of software dataset by statistically identifying the attributes and projects that satisfy EBA hypothesis. The proposed method was motivated by the challenges in ANGEL and Analogy-X in addition to the need to improve reliability of EBA method especially when using too many categorical attributes. The resulting method forms as an extension for EBA method which resulted in a new software effort estimation method called EBA$^+$.

EBA$^+$ was evaluated against ANGEL **[18, 19]**, and Analogy-X **[16]**, using normalized Euclidian distance and closest analogy. Moreover, six datasets have been used for empirical evaluations: ISBSG **[12]**, Desharnais **[7, 8]**, COCOMO **[6, 7]**, Kemerer **[7, 14]**, Maxwell **[7]** and Albrecht **[1, 7]**. The main evaluation results are: (1) the proposed method could form a useful extension for EBA; (2) It is able to identify whether the categorical attribute is indeed appropriate for making prediction or not. (3) It is able to identify reliable attributes and projects through robust statistical procedure.

The rest of the paper is subdivided into 8 sections: section 2 presents overview of EBA method and its assumptions. Section 3 introduces Kendall row-wise correlation and its significance test. Section 4 introduces EBA$^+$ framework. Section 5 presents evaluation criteria used to validate EBA$^+$. Section 6 presents the results obtained from empirical evaluation. Section 7 presents discussion and conclusions.

## 2. Estimation by analogy and its fundamental assumption

EBA is one of the important estimation methods in software engineering domain which aims to identify solution for a new problem based on previous solutions from the set of similar cases **[18, 19, 24]**. The general process of basic EBA method as shown in Figure 1 consists of four main stages:
1. Dataset preparation including attribute selection and handling missing values.
2. Defining a new project that will be estimated.
3. Retrieving the analogues of the new project using a predefined similarity function. The common used function is the Euclidean distance function as depicted in Eq. (1).

$$d(p_i, p_j) = \frac{1}{m}\sqrt{\sum_{k=1}^{m} \Delta(p_{ik}, p_{jk})} \quad (1)$$

where d is the Euclidean distance, m is the number of predictor attributes, $P_i$ and $P_j$ are projects under investigation and:

$$\Delta(p_{ik}, p_{jk}) = \begin{cases} \frac{(p_{ik} - p_{jk})^2}{\max_k - \min_k} & \text{if feature k is continuous} \\ 0 & \text{if feature k is categorical and } p_{ik} = p_{jk} \\ 1 & \text{if feature k is categorical and } p_{ik} \neq p_{jk} \end{cases} \quad (2)$$

Where value $\max_k$ and $\min_k$ are the maximum and minimum values of attribute k respectively.

4. Predicting effort of the new project from the retrieved analogues.

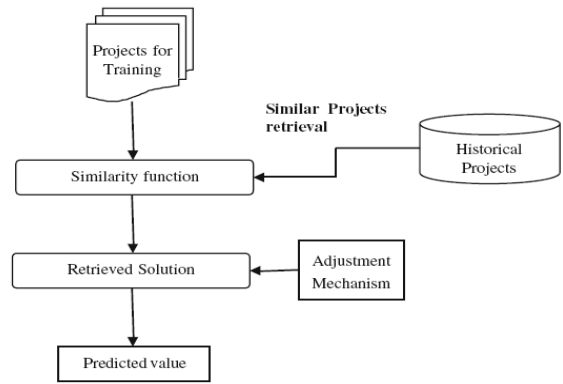

Figure 1. Process of analogy based estimation method

The process of EBA method is mainly performed based on the following fundamental assumption: *"The projects that are similar in terms of their attributes values will also be similar in terms of their effort values"* **[24].** The aforementioned assumption has two major issues. The first issue implies that any attribute is regarded as useful for prediction only if it has a strong relationship with the effort, otherwise it is irrelevant **[2, 4, 20, 21]**. The second issue implies that for any target project the ranks of similar source projects with respect to a specific attribute should be consistent with ranks of similar source projects with respect to their effort values. Both issues can be tested using Kendall's row-wise rank correlation **[25]** between similarity matrix based project attribute values and similarity matrix based effort values. In summary, this paper presents a new method to: (1) Test the appropriateness of datasets for EBA method. (2) Identify reliable attributes and (3) Identify abnormal projects that undermine estimation process.

## 3. Kendall row-wise correlation
### 3.1 Overview

Kendall Rank Correlation was developed by Maurice Kendall in 1938 **[15]** to measure the degree of correspondence between two sets of ranks given to a same set of objects and assessing the significance of this correspondence. Kendall rank correlation coefficient depends upon the number of inversions of pair of objects which would be required to transform one rank order into another. Kendall rank coefficient of correlation can take values between -1 and +1 corresponding to the range from perfect strong disagreement to perfect strong agreement. If the rankings are completely independent, the Kendall rank coefficient is 0.

For the sake of evaluating EBA assumption we used the alternative method of Kendall rank correlation which is the Kendall rowwise rank correlation (CORR) that measures correlation between two similarity matrices **[25]** as shown in Eq. (4). This kind of correlation is based upon a weighted sum of the correlations between all pairs of corresponding rows of two similarity matrices, bearing in mind that the diagonal elements are not considered in the calculation. For ranking elements in each individual row, we used mid-ranking technique ("1 2.5 2.5 4") that assigns objects that have equal value the same ranking number, which is the average of what they would possess under ordinal rankings. Figure 2 shows a typical example of two similarity matrices SM(X) and SM(Y) where $X_{12}$ represents similarity degree between project $P_1$ and $P_2$ in terms of attribute X. Interestingly, some researchers refer to a similarity matrix (where 1 means two different projects are equivalent with respect to all project factors) and others to a distance matrix (where 1 means two projects are completely different with respect to all project factors). Indeed, however, elements of a similarity matrix are just 1-elements (or reciprocal) of a normalized distance matrix (i.e. Similarity Matrix = 1 – Normalized Distance Matrix)

$$SM(X) = \begin{bmatrix} 1 & X_{12} & \cdots & X_{1j} & X_{1k} & \cdots X_{1n} \\ X_{21} & 1 & \cdots & X_{2j} & X_{2k} & \cdots X_{2n} \\ \vdots & \vdots & 1 & \vdots & \vdots & \vdots \\ X_{i1} & X_{i2} & \cdots & 1 & X_{ik} & \cdots X_{in} \\ \vdots & \vdots & \cdots & \vdots & \vdots & \vdots \\ X_{n1} & X_{n2} & \cdots & X_{nj} & X_{nk} & \cdots 1 \end{bmatrix} \quad SM(Y) = \begin{bmatrix} 1 & Y_{12} & \cdots & Y_{1j} & Y_{1k} & \cdots Y_{1n} \\ Y_{21} & 1 & \cdots & Y_{2j} & Y_{2k} & \cdots Y_{2n} \\ \vdots & \vdots & 1 & \vdots & \vdots & \vdots \\ Y_{i1} & Y_{i2} & \cdots & 1 & Y_{ik} & \cdots Y_{in} \\ \vdots & \vdots & \cdots & \vdots & \vdots & \vdots \\ Y_{n1} & Y_{n2} & \cdots & Y_{nj} & Y_{nk} & \cdots 1 \end{bmatrix}$$

**Figure 2** General form of two similarity matrices

$$\text{CORR} = \frac{\sum_{i,j,k} \text{sign}(X_{ij} - X_{ik}) \text{sign}(Y_{ij} - Y_{ik})}{\left( \sum_{i,j,k} \text{sign}(X_{ij} - X_{ik})^2 \sum_{i,j,k} \text{sign}(Y_{ij} - Y_{ik})^2 \right)^{1/2}}, \tag{4}$$

where $i = 1, \cdots, n; \; j, k = 1, \cdots, n \;; \; j < k; \; i \neq k, \; i \neq j$

$$\text{and sign}(X_{ij} - X_{ik}) = \begin{cases} +1 & X_{ij} > X_{ik} \\ 0 & X_{ij} = X_{ik} \\ -1 & X_{ij} < X_{ik} \end{cases} \tag{5}$$

For example, suppose that there are two similarity matrices SM(X) and SM(Y) using 5 hypothetical projects, as shown in Figure 3. Applying Kendall's row-wise rank correlation on (SM(X) vs. SM(Y)) resulted in CORR(X) =+0.8667 which indicates strong agreement between attributes X and Y. It is important to note that we use CORR(X) for short to represent Kendall row-wise correlation value between similarity matrix based attribute X and similarity matrix based predictable attribute Y.

SM(X)

|       | $p_1$ | $p_2$ | $p_3$ | $p_4$ | $p_5$ |
|-------|------|------|------|------|------|
| $p_1$ | 1    | 0.8  | 0.53 | 0.67 | 0.47 |
| $p_2$ | 0.8  | 1    | 0.33 | 0.47 | 0.67 |
| $p_3$ | 0.53 | 0.33 | 1    | 0.87 | 0    |
| $p_4$ | 0.67 | 0.47 | 0.87 | 1    | 0.13 |
| $p_5$ | 0.47 | 0.67 | 0    | 0.13 | 1    |

SM(Y)

|       | $p_1$ | $p_2$ | $p_3$ | $p_4$ | $p_5$ |
|-------|------|------|------|------|------|
| $p_1$ | 1    | 0.93 | 0.31 | 0.60 | 0.69 |
| $p_2$ | 0.93 | 1    | 0.24 | 0.53 | 0.76 |
| $p_3$ | 0.31 | 0.24 | 1    | 0.71 | 0    |
| $p_4$ | 0.60 | 0.53 | 0.71 | 1    | 0.29 |
| $p_5$ | 0.69 | 0.76 | 0    | 0.29 | 1    |

Row-wise ranks of SM(X)

|       | $p_1$ | $p_2$ | $p_3$ | $p_4$ | $p_5$ |
|-------|------|------|------|------|------|
| $p_1$ | *    | 1    | 3    | 2    | 4    |
| $p_2$ | 1    | *    | 4    | 3    | 2    |
| $p_3$ | 2    | 3    | *    | 1    | 4    |
| $p_4$ | 2    | 3    | 1    | *    | 4    |
| $p_5$ | 2    | 1    | 4    | 3    | *    |

Row-wise ranks of SM(Y)

|       | $p_1$ | $p_2$ | $p_3$ | $p_4$ | $p_5$ |
|-------|------|------|------|------|------|
| $p_1$ | *    | 1    | 4    | 3    | 2    |
| $p_2$ | 1    | *    | 4    | 3    | 2    |
| $p_3$ | 2    | 3    | *    | 1    | 4    |
| $p_4$ | 2    | 3    | 1    | *    | 4    |
| $p_5$ | 2    | 1    | 4    | 3    | *    |

**Figure 3.** SM(X) vs. SM(Y) and their row-wise ranking orders

### 3.2 Kendall's rowwise rank correlation for nominal data

Software projects are not only described by numerical attributes, but often with categorical (ordinal and nominal) attributes as well. The way to assess similarity degree for nominal attribute values is kind of comparison: 1 when they are equivalent and 0 otherwise. However, we have seen earlier that Kendall rowwise correlation can work well with numerical attributes, but a common problem occurs when one of the similarity matrices provides a ranking and the other a dichotomy or classification into two classes (like similarity degree between nominal attribute values). In the introduction we mentioned that Analogy-X method cannot directly deal with the distance matrix for nominal data individually. So it cannot assess the appropriateness of nominal attribute individually, but alternatively it can test the impact of nominal attribute by dividing the selected continuous attributes into subgroups according to the number of categories in that nominal attribute. This solution removes large number of elements in the matrix, because only matrix elements within the same category are considered in the correlation calculation. To overcome this challenge, the Kendall rowwise correlation has ability to test the quality of nominal attribute solely without the need to partition the dataset into subgroups. To illustrate how Kendall rank correlation handles this problem, consider the following similarity degrees between different source projects and target project $P_t$ based on nominal attribute Z and predictable attribute Y.

**Table 1** Similarity degrees between target project $p_t$ and other source projects

| Source project | $p_1$ | $p_2$ | $p_3$ | $p_4$ | $p_5$ | $p_6$ | $p_7$ | $p_8$ |
|---|---|---|---|---|---|---|---|---|
| Z | 1 | 0 | 1 | 0 | 1 | 1 | 0 | 1 |
| Y | 0.5 | 0.3 | 0.7 | 0.0 | 0.9 | 0.1 | 0.4 | 0.6 |

Kendall's correlation assumes that the division into binary is itself ranking. For example, in Table 1 there are 5 ones and 3 zeros for which Kendall's correlation supposes that the first 5 members of 1's are tied, and also for the next group of 0's. According to mid-ranking method, the average of the tied ranks will be 3 in the first group of ties and 7 in the second group of ties, so that the pair of ranking appeared as shown in Table 2. Accordingly, the CORR(Z) is +0.504 which indicates a degree of agreement between their rankings.

**Table 2** Ranks of similarity degrees in Table 3

| Source project | $p_1$ | $p_2$ | $p_3$ | $p_4$ | $p_5$ | $p_6$ | $p_7$ | $p_8$ |
|---|---|---|---|---|---|---|---|---|
| Z | 3 | 7 | 3 | 7 | 3 | 3 | 7 | 3 |
| Y | 4 | 6 | 2 | 8 | 1 | 7 | 5 | 3 |

In the following example (shown in Figure 4) we want to investigate the degree of association between similarity matrix based nominal attribute SM(Z) and similarity matrix based predictable attribute SM(Y). The resulted CORR(Z) is 0 which confirms that the nominal attribute Z is not strongly correlated with dependent attribute E. This implies that the nominal attribute Z is not useful if it is taken individually for identifying a similar project, but its effect could be observed if it is combined with continuous or ordinal attributes in the similarity matrix.

```
              SM(Z)                              SM(Y)

       p1  p2  p3  p4  p5                 p1    p2    p3    p4    p5
  p1   1   1   0   0   0           p1     1    0.93  0.31  0.60  0.69
  p2   1   1   0   0   0           p2   0.93    1    0.24  0.53  0.76
  p3   0   0   1   0   1           p3   0.31  0.24    1    0.71    0
  p4   0   0   0   1   0           p4   0.60  0.53  0.71    1    0.29
  p5   0   0   1   0   1           p5   0.69  0.76    0    0.29    1
```

**Figure 4** Kendall's rowwise correlation between SM(Z) vs. SM(Y)

### 3.3 Significance test for Kendall row-wise rank correlation

The significance test of Kendall rowwise correlation is a primary source of information about the reliability of the correlation and therefore the reliability of the dataset for EBA method. It is usually tested by using a permutation test, at a significance level of 5% **[25]**. The original correlation value is compared with the correlation values found by randomly generating a set of permutations of the rows (and simultaneously of the columns) for one of the two matrices, and calculating the value of correlation (CORR) for each permutation. The significance of the observed value of the statistic is then assessed by calculating the proportion of values as large as or larger than original correlation value, i.e. right tail probability **[25]**.

The reason for this procedure is that if the null hypothesis of no correlation between two matrices cannot be rejected, then permuting the rows and columns of the matrix should be equally likely to produce a larger or similar correlation coefficient. However, using this permutation we can test whether the value of Kendall's row-wise rank correlation derived from the original pair of similarity matrices is significantly different from zero. If so, then we can be sure that the fundamental assumption of EBA is true, which consequentially suggests using that attribute(s). The required number of permutations to accomplish significance test is not limited irrespective of existing some recommendations. The default configuration recommends 1000 permutations for estimating the significance level of about 0.05 and 5,000 permutations for estimating significance level of about 0.01 **[22]**.

## 4. The EBA$^+$ method

EBA$^+$ framework utilizes Kendall row-wise rank correlation and its significance test to determine the appropriateness of the dataset under consideration, which implicitly includes selecting reliable attributes and removing abnormal projects. The EBA$^+$ framework will test the following hypothesis:
**H0** (null hypothesis): The dataset is not reliable for EBA.
**H1** (alternative hypothesis): The dataset is reliable for EBA.

The EBA$^+$ framework cannot reject null hypothesis in the following two cases: (1) when no attribute satisfies hypothesis of EBA, and (2) when all projects are abnormal. Otherwise the null hypothesis is rejected and the dataset is reliable for EBA. The work flow of this process can be further illustrated in Figure 5. To integrate abnormal project identification process with attribute selection process, we suggest first performing the attribute selection and then performing abnormal project identification process. If significantly abnormal projects are detected remove them and update the reduced dataset.

The application of EBA$^+$ method requires a robust measure to estimate the precision of correlation coefficient for each attribute. Therefore we used Bootstrapping method [9] that draws samples randomly with replacement from the empirical distribution of data to estimate the shape of a statistic's sampling distribution. Bootstrapping is a non-parametric statistical method for deriving confidence intervals and estimates of standard errors for different estimates such as the correlation coefficient, regression coefficient, mean, median or proportion [9]. One issue with the bootstrapping method is that the theory relies on datasets being random samples from a well-defined population. This assumption is probably not true for software project datasets but it probably will not make much practical difference. The training function used in our case is the Kendall rowwise correlation (CORR) for the drawn samples as shown in Eq. (6). Given a distribution of n samples from Bootstrapping training, the overall Kendall correlation value ($\tau_r$) is calculated as shown in Eq. (6). The $\tau_r$ will be used thereafter to represent Kendall rowwise correlation instead of CORR.

$$\tau_r = \frac{1}{n}\sum_i \text{CORR}(i) \qquad (6)$$

The corresponding upper and lower confidence interval limit (UCL and LCL) of Kendall correlation coefficient can be approximated by using BCa method (BCa stands for Bias-corrected and accelerated). BCa **[9]** is a reliable method to derive confidence interval especially when the distributions are skewed. The UCL and LCL is calculated using Eqs. (7) and (8).

$$\text{LCL} = \Phi\left(z_o + \frac{z_o + z^{0.025}}{1 - a(z_o + z^{0.025})}\right) \qquad (7)$$

$$\text{UCL} = \Phi\left(z_o + \frac{z_o + z^{0.975}}{1 - a(z_o + z^{0.975})}\right) \qquad (8)$$

Where $z_o$ is the bias-correcting constant, which is the standard normal deviate corresponding to the proportion of bootstrap estimates which are less than or equal to the estimate from the original sample. $\Phi$ represents the cumulative distribution of the standard normal function. $\alpha$ is the desired level of significance eg. 0.05. $z^{\alpha}$ is the $\alpha$-percentile of a standard normal distribution.

The EBA⁺ method as shown in Figure 5 consists of three stages as follows:

**Stage 1**: The process of EBA⁺ starts with testing every attribute individually. For each attribute X, the Kendall row-wise correlation between SM(X) and SM(Effort) is computed. Then all attributes that their similarity matrices are significantly correlated with similarity matrix of project effort are selected. If no one attribute is selected then the dataset is not entirely reliable and does not satisfy the hypothesis of EBA. In this case the algorithm should terminate and no need to continue. If only one attribute is selected then this attribute is considered as the reliable one, then move to stage 3. Otherwise, if the number of attribute is greater than one then move to stage 2.

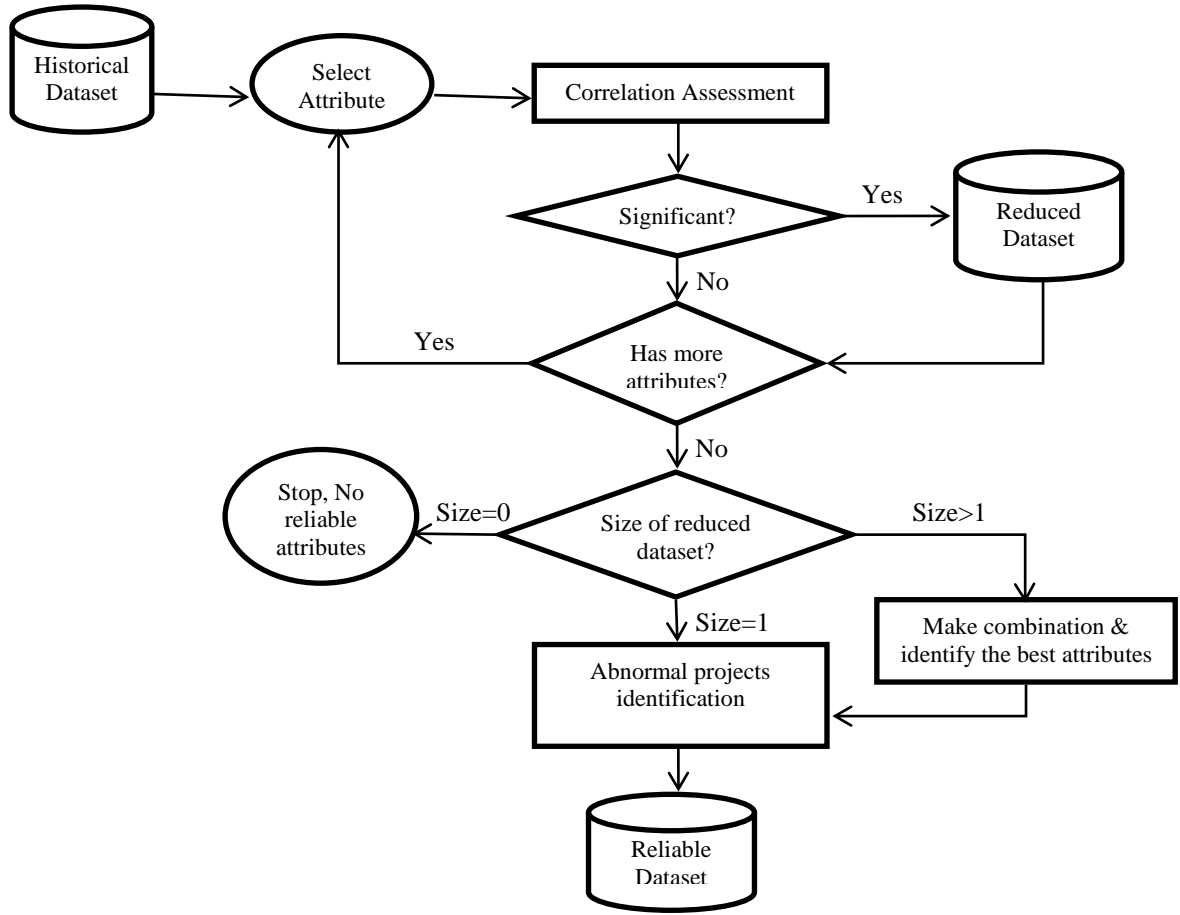

**Figure 5** illustration of EBA⁺

**Stage 2**:
1. Identify the best significant attribute from stage 1 for which $\tau_r$ is greatest (say Best).
2. For the remaining selected attributes, calculate a similarity matrix for each attribute in combination with Best. If the best Kendall rowwise correlation obtained is less than or equal to $\tau_r$(Best) and CI is wider than CI(Best), stop and use Best alone for EBA.
3. If the best Kendall rowwise correlation obtained is greater than $\tau_r$(Best) and CI is narrower than CI(Best), choose that attribute and call that new set of attributes Y. The justification for that having a narrow confidence interval implies high precision whilst a wide interval implies poor precision.
4. Steps 2 and 3 are repeated by adding one attribute at a time, until maximum Kendall correlation and narrower CI are obtained. The produced dataset from this stage is called Reduced Dataset (RD1).

**Stage 3**: The aim of stage is to detect the abnormal projects that do not satisfy the hypothesis of EBA in RD1. Once the abnormal projects are identified they are removed from the reduced dataset. To perform this stage, the similarity matrix of RD1 is constructed first; say SM(RD1) in which each row contains similarity degrees between a target project and other source projects. Then the significance of correlation for each pair of the corresponding rows in both SM(RD1) and SM(Effort) is computed. If the correlation is insignificant then the target project of that row is removed from RD1. This procedure is performed for all projects in RD1. To better understand the EBA[+] we provide the pseudo code in Figure 6.

```
// input: Dataset D
// output: ReducedDataset RD
nprem=1000         //Number of permutations
nSample=1000       //number of bootstrapping samples
nProjects          //number of projects in the dataset
SM(E)=FindSim(Effort) // construct similarity matrix of effort values
All=[];
//stage 1: Find individual significant attributes
Foreach attribute X in the dataset D
      FOR i=1 to nSamples
         R_i← Draw(X)//Draw a resample with replacement from the attribute X.
         SM(R_i)←FindSimMatrix(R_i)  //construct similarity matrix of X
         CORR[i]←KendalRowWise(SM(R_i),SM(E))
      END
      τ_r(X)←mean(CORR)
      CI(X)←Distribution(CORR);//compute conf. interval of X (UCL- LCL)
      pval(X)←permute(SM(X), SM(E),τ_r(X), nPrem)
         IF(pval(X)<0.05)
            All←All+X;
         END
END
All← Sort(All);//sort all the significant attributes according to their
correlation values.
Best← All[1];

//Stage 2: Find the reliable attribute combination
For j=2 to size(All)
      Y=Best+ All[j]
      SM(Y)←FindSimMatrix(Y);
      FOR i=1 to nSamples
         R_i=Draw(Y)//Draw a resample with replacement from the attribute Y.
         SM(R_i)=FindSim(R_i)  // construct similarity matrix of Y
         CORR[i]=KendalRowWise(SM(R_i),SM(E))
      END
      τ_r(Y)=mean(CORR)
      CI(Y)←Distribution(CORR);//compute conf. interval of Y (UCL- LCL)
         If(( CI(Y) > CI(X)) || τ_r(Y) < τ_r(Best))
            Y←Y-All[j]
         END
         Best=Y
End
RD1=Best

//stage 3: Abnormal project detection
SM(RD1)←FindSimMatrix(RD1);
For idx=1 to nProjects
  Corr←KendalRowWise(SM(RD1)[idx], SM(E)[idx])
  pvalue←permute(SM(RD1)[idx], SM(E)[idx], Corr, nPrem)
  IF(pvalue<0.05)
     RD←RD1[idx]
  End
End
```

Figure 6. Pseudo Code of EBA$^+$

## 5. Performance measures

Three well known evaluation criteria were used to assess the degree of accuracy to which the estimated efforts match actual efforts:

(i) Magnitude Relative Error (MRE), as depicted in Eq. (9), computes the absolute percentage of error between actual and predicted effort for each reference project.

$$\text{MRE}_i = \frac{|\text{actual}_i - \text{estimated}_i|}{\text{actual}_i} \tag{9}$$

(ii) Mean magnitude relative error (MMRE), as shown in Eq. (10), calculates the average of MRE over all reference projects. Since the MMRE is sensitive to an individual outlying prediction, when we have a large number of observations, we adopt median of MREs for the n projects (MdMRE) as shown in Eq. (11), which is less sensitive to the extreme values of MRE. However, MMRE is not always reliable to compare prediction methods because it has been criticized that is unbalanced in many validation circumstances and leads often to overestimation **[11]**. Therefore we used Wilcoxon sum rank significance test to compare between the median of two samples based on absolute residuals, setting the confidence limit at 0.05.

$$\text{MMRE} = \frac{1}{n}\sum_{i=1}^{n}\text{MRE}_i \tag{10}$$

$$\text{MdMRE} = \underset{i}{\text{median}}(\text{MRE}_i) \tag{11}$$

(iii) PRED(25) as depicted in Eq. (12) is used to count the percentage of estimates that have MRE less than 25%.

$$\text{PRED}(25) = \frac{\lambda}{n} * 100 \tag{12}$$

where $\lambda$ is the number of projects where $\text{MRE}_i \leq 25\%$, and n is the number of all observations. A software estimation model with lower MMRE, MdMRE, and higher PRED(25) shows that its derived estimates are more accurate than other models.

## 6. Experimental Results

### 6.1 Empirical evaluation of EBA$^+$

This section empirically examines the performance of the EBA$^+$ algorithm against ANGEL's Brute-force algorithm (ANGEL for short) **[18]** and Analogy-X **[16]** through a series of evaluation studies with various datasets that exhibit typical characteristics of software effort estimation. Thus, we believe that consistent results obtained using all these datasets validate our algorithm. These datasets come from different sources: **Maxwell** [7], ISBSG release 10 **[12]**, Desharnais **[7]**, Kemerer **[7]**, Albrecht **[7]** and COCOMO'81 **[7]**.

For the sake of validation and comparison, in this section, we will use normalized Euclidean distance as similarity measure and only the closest analogy. To normalize the Euclidean distance each distance was divided by the maximum of distance obtained for the whole matrix. The resulting normalized Euclidean distances are in range of 0 to 1. Then the normalized similarity degree is 1-normalized distance degree. Furthermore, for the purpose of prediction accuracy comparison between reduced model of EBA$^+$ and other models we used Jackknife validation procedure. In Jackknifing validation, one observation is held out once as test data and the model is trained on the remaining observations, then its MRE is evaluated. Thus, the evaluation procedure is executed *n* times according to the number of observations **[5, 23]**.

## 6.2 Maxwell dataset

This section presents an example illustrating the process of EBA$^+$ on Maxwell dataset **[7]**. The Maxwell dataset is a relatively new dataset, which contains 62 projects described by 23 features, collected from one of the biggest commercial banks in Finland. The dataset includes larger proportion of categorical features (with 22 features) which is fair to demonstrate the performance of EBA$^+$. The full description of Maxwell dataset is presented in Table 3. The only numerical attribute is the project Size in Function points.

**Table 3** Maxwell dataset description

| Feature | Description | $\tau_r$ | p-value | LCL | UCL |
|---|---|---|---|---|---|
| App | Application type | -0.0152 | 0.80 | -0.048 | 0.017 |
| Har | Hardware platform | 0.0364 | 0.05 | 0.0037 | 0.062 |
| Dba | Database | 0.0943 | 0.04* | 0.059 | 0.126 |
| Ifc | User interface | -0.0371 | 0.91 | -0.0424 | -0.029 |
| Source | Where developed | -0.0401 | 0.85 | -0.0827 | -0.007 |
| Telonuse | Telon use | -0.005 | 0.54 | -0.031 | 0.0154 |
| Nlan | # of development languages | 0.0305 | 0.03* | -0.0124 | 0.0738 |
| T01 | Customer participation | 0.0461 | 0.01* | 0.0017 | 0.0878 |
| T02 | Development Env. adequacy | 0..0511 | 0.02* | 0.0247 | 0.0744 |
| T03 | Staff availability | -0.0117 | 0.69 | -0.0404 | 0.011 |
| T04 | Standards use | -0.019 | 0.87 | -0.0306 | -0.0082 |
| T05 | Methods use | 0.0819 | 0.02* | 0.0297 | 0.124 |
| T06 | Tools use | 0.0315 | 0.08 | 0.0035 | 0.0524 |
| T07 | Software logical complexity | 0.097 | 0.01* | 0.05 | 0.149 |
| T08 | Requirements volatility | 0.0126 | 0.12 | -0.0202 | 0.0423 |
| T09 | Quality requirements | -0.0023 | 0.57 | -0.0422 | 0.029 |
| T10 | Efficiency requirements | 0.0236 | 0.04* | -0.0217 | 0.0624 |
| T11 | Installation requirements | 0.0379 | 0.01* | 0.0057 | 0.0675 |
| T12 | Staff analysis skills | -0.0159 | 0.7 | -0.0532 | 0.0184 |
| T13 | Staff application knowledge | -0.0315 | 0.97 | -0.0565 | -0.009 |
| T14 | Staff tool skills | 0.0457 | 0.01* | 0.0174 | 0.071 |
| T15 | Staff team skills | 0.0629 | 0.02* | 0.0184 | 0.0958 |
| Size | Function points | 0.3309 | 0.01* | 0.292 | 0.367 |
| *: correlation is significant at 0.05 | | | | | |

Applying stage 1 of EBA$^+$ method by computing $\tau_r$ between each SM($X_i$) and SM(effort) indicates that 11 attributes were significantly correlated with the effort (i.e. p-value < 0.05) as shown in Table 3. These attributes are: Dba, Nlan, T01, T02, T05, T07, T10, T11, T14, T15 and Size. The strong correlated similarity matrix was obtained by Size attribute with Bootstrap estimator $\tau_r$(Size)= +0.3309 and CI=[0.292 0.367], which indicates a positive agreement between Size and effort. At this stage we cannot consider that all identified attributes are useful for EBA until we ensure that there is no attribute(s) in combination with Size attribute produce(s) larger correlation coefficient and narrower CI than that of Size alone. Since the number of selected attributes is greater than one, stage 2 of EBA$^+$ starts execution by recursively adding one attribute a time to the best attribute(s) and checks their correlation coefficients. However, It was found that the similarity matrix of Size and Nlan attributes together produced correlation coefficient value larger than $\tau_r$(Size) with Bootstrap estimator of correlation $\tau_r$(Size and Nlan) = + 0.342, and a narrower confidence interval [0.304 0.368]. The abnormal project identification process revealed that the three projects (P4, P12 and P20[1]) are considered significantly abnormal with p-value > 0.05. The results obtained confirm that attributes Size and Nlan are still the most influential attributes with an improved correlation coefficient $\tau_r$(Size and Nlan) =0.344. No further abnormal project cases exist in the dataset which in this case, the obtained data from this combination are the most influential attributes and will be used to generate new estimates. However, the null hypothesis is rejected and the reduced dataset is reliable for EBA method.

---

[1] According to dataset available at PROMISE website

To investigate the effectiveness of the reduced dataset on the prediction accuracy, we made a comparison against attributes identified by ANGEL and Analogy-X. For the sake of validation we used only the closest analogy. Running ANGEL and Analogy-X yielded different attribute combinations in that the ANGEL identified the same attributes of EBA$^+$, while Analogy-X identified only one attribute which is (Size). The significant difference between EBA$^+$ and ANGEL is the ability of EBA$^+$ to identify abnormal projects. In other words, the attributes identified by ANGEL are not necessarily reliable as much as they are only predictive. Unlike ANGEL which continues to generate predictions regardless of reliability of the dataset, EBA$^+$ and Analogy-X attempt to search for the robust attributes that have sound statistical significance. In addition, if one attempts to find the best attributes based on optimizing PRED(25), it may obtain the best PRED(25) but not always the best MMRE and vice versa. This explains why relying on MMRE or PRED(25) could not be the best solution to search for best attributes as the prediction accuracy can be measured by various performance indicators. Therefore it is reasonable to believe that the combination of attributes Size and Nlan will be useful for EBA regardless of whichever performance indicator is used.

However, the comparison of prediction accuracy based upon using EBA$^+$, ANGEL, and Analogy-X methods are shown in Table 4. The predictive performance results show that the EBA$^+$ produced better accuracy than other methods in terms of all performance indicators.

**Table 4** Prediction accuracy comparison for Maxwell dataset

| Algorithm | Best Attribute set | MMRE | PRED(25)% | MdMRE |
|---|---|---|---|---|
| ALL | ALL | 182.0 | 9.6 | 72.25 |
| EBA$^+$ | Size and Nlan (with three projects removed P4, P12 and P20) | **51.64** | **25.4** | **38.2** |
| ANGEL | Size and Nlan | 65.83 | 20.96 | 40.93 |
| Analogy-X | Size | 90.7 | 22.6 | 58.2 |

The Boxplot of absolute residuals in Figure 7 and Wilcoxon sum rank significance test indicate that the difference between the predictions generated based on attributes identified by EBA$^+$ and those based on Analogy-X are statistically significant (p-value=0.02). This means that there is difference between their predictions. In contrast, the predictions generated by EBA$^+$ and ANGEL are insignificant (p-value=0.32) which means that the predictions generated by both methods are approximately similar. The insignificant results confirm that the EBA$^+$ method can find the best predictive attributes and projects based on external measure (i.e. Kendall correlation) not like ANGEL which is optimized based on internal measure(i.e. MMRE). This comes to conclude that EBA$^+$ has the ability to identify significantly, with justification, the reliable data that produces comparable but necessarily reliable results.

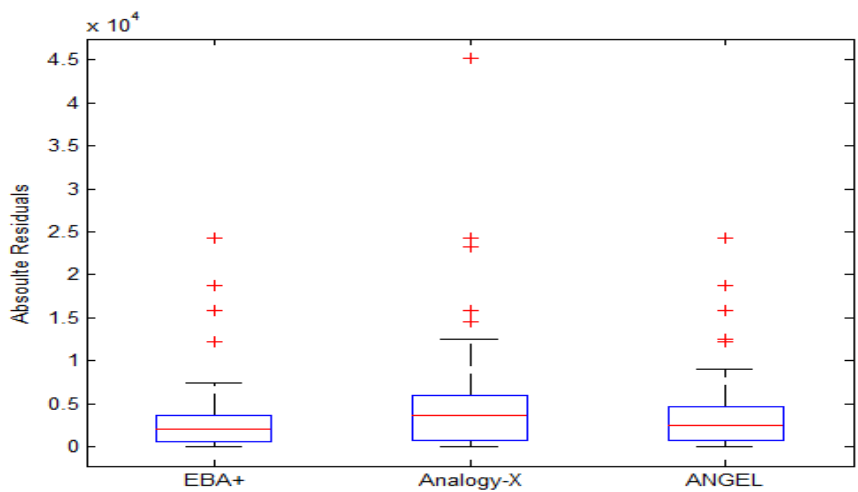

**Figure 7** Box-plot of absolute residuals of prediction using EBA$^+$, Analogy-X, and ANGEL for Maxwell dataset

## 6.3 Albrecht dataset

The Albrecht dataset contains 24 software projects developed using third generation languages such as COBOL, PL1, etc. The dataset is described by one dependent attribute called 'work hours' which represents the corresponding effort in 1000 hours, and six independent numeric attributes as shown in Table 5. 18 projects were written in COBOL, 4 projects were written in PL1 and the rest were written in database management languages. Two projects have effort values more than 100,000 hours which are twice larger than third largest project. These extreme projects have considerable negative impact on prediction but we preferred to keep them in spite of their potential bad consequences.

Table 5 Albrecht dataset description

| Attribute | Description | $\tau_r$ | $p$-value | LCL | UCL |
|---|---|---|---|---|---|
| *KLOC* | Line of code | +0.43* | 0.001 | 0.38 | 0.490 |
| *AFP* | Adjusted Function Points | +0.52* | 0.001 | 0.44 | 0.584 |
| *INC* | Count of input functions | +0.182* | 0.001 | 0.086 | 0.243 |
| *OC* | Count of output functions | +0.496* | 0.001 | 0.410 | 0.552 |
| *INQ* | Count of query functions | +0.142* | 0.006 | 0.086 | 0.195 |
| *FC* | Count of file processing | +0.236* | 0.001 | 0.179 | 0.284 |
| *RawFP* | Raw Function Points | +0.49* | 0.001 | 0.423 | 0.533 |
| *: correlation is significant at 0.05 | | | | | |

The results of stage 1 of EBA[+] indicate that all attribute similarity matrices were significantly correlated with the effort similarity matrix (i.e. p-value < 0.05) as shown in Table 5. The strong correlated similarity matrix from stage 1 was obtained by AFP with Bootstrap estimator $\tau_r(\text{AFP}) = +0.52$ and CI=[0.44  0.584] which indicates a strong positive agreement between AFP and effort. The stage 2 found that the similarity matrix of (AFP, INQ and KSLOC) attribute together produced correlation coefficient value larger than $\tau_r(\text{AFP})$ with Bootstrap estimator of correlation $\tau_r(\text{AFP} + \text{INQ} + \text{KSLOC}) = +0.53$ and a narrower confidence interval. The abnormal project identification process revealed that the two projects (P5 and P21) are considered significantly abnormal with p-value > 0.05. The results obtained confirm that attributes AFP, INQ and KSLOC are still the most influential attributes, with an improved correlation coefficient $\tau_r(\text{AFP} + \text{INQ} + \text{KSLOC}) = +0.54$. No further abnormal project cases exist in the dataset which in this case, the obtained attributes from this combination are the most influential attributes and will be used to generate new estimates. The obtained results thus reject the null hypothesis and consider that the dataset is reliable for EBA.

However, ANGEL identified four predictive attributes (INQ, FC, KSLOC, AFP), while Analogy-X identified only one attribute as predictive attribute which is (RawFP). No projects have been removed by both algorithms. Interestingly, there are two attributes that are common between EBA[+] and ANGEL which shows the importance of these two attributes. In contrast, there is no any attribute in common between Analogy-X and others. Comparisons of prediction accuracy based upon using EBA[+] and other (ANGEL, Analogy-X) attributes selection approaches are shown in Table 6. The results obtained demonstrate that predictions generated using EBA[+] are nearly more accurate than others in terms of MMRE and PRED(25). Also the attributes identified by ANGEL produced better MMRE value than Analogy-X that produced worst predictions and even worse than using all attributes with MMRE=97.8%. The main reason may be that ANGEL attempts to search for attribute based on optimizing MMRE, therefore the MMRE obtained by ANGEL is regarded as the best value can be achieved.

Table 6 Prediction accuracy comparison for Albrecht dataset

| Algorithm | Best Attribute set | MMRE | PRED(25)% | MdMRE |
|---|---|---|---|---|
| ALL | ALL | 71.0 | 29.16 | 38.9 |
| EBA[+] | *AFP+INQ+KSLOC* (with two projects removed (P5 and P21)) | **59.2** | **41.67** | 33.2 |
| ANGEL | *INQ+FC+KSLOC+AFP* | 63.5 | 33.33 | 38.9 |
| Analogy-X | *RawFP* | 97.8 | 44.1 | **29.17** |

Since MMRE is not widely recommended to compare between prediction models **[14]**, we used Boxplots of absolute residuals and Wilcoxon sum rank test to investigate whether the differences in absolute residuals of predictions are statistically significant.

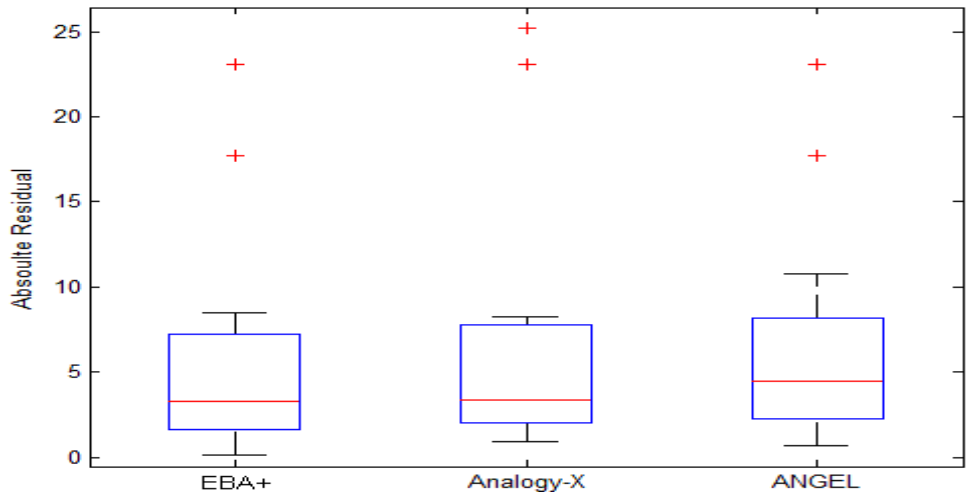

**Figure 8** Box-plot of absolute residuals of prediction using EBA+, Analogy-X, and ANGEL for Albrecht dataset

**Table 7** Wilcoxon sum rank test for Albrecht dataset

| Algorithms | p-value | Rank sum |
|---|---|---|
| EBA+ Vs. ANGEL | 0.783 | 576 |
| EBA+ Vs. Analogy-X | 0.881 | 582.5 |

The results obtained in Table 7 and Figure 8 indicate that the difference between predictions generated based on attributes identified by EBA+ and those based on ANGEL and Analogy-X are statistically insignificant (p-value > 0.05). Furthermore, there is sufficient evidence that all predictions using different combinations of attributes are approximately similar. This comes to conclude that EBA+ has the ability to identify significantly, with justification, the optimal attribute subset that produces comparable accuracy.

### 6.3.1 Kemerer dataset

The Kemerer dataset **[7]** includes 15 software projects described by 6 attributes and one predictable attribute which is measured in 'man-month'. The 6 attributes are represented by 2 categorical and 4 numerical attributes.
A similar procedure to the previous case study was applied to Kemerer dataset. The EBA+ identified 4 significant correlated similarity matrices based attributes: KSLOC, AdjFP, RawFP and Language individually. This confirms that the dataset has reliable attributes that satisfy hypothesis of EBA. The strong significant correlated similarity matrix was obtained by attribute AdjFP with $\tau_r(\text{AdjFP}) = +0.336$ and CI=[0.1612  0.4491]. It is interesting to note that the categorical attribute Language was also significant with p-value < 0.05 which indicates that the EBA+ is able to identify whether the categorical attribute is reliable for predictions or not. In the final step of EBA+ stage 2 we found that the similarity matrix based on combination of AFP and Language was stronger correlated than best correlated attribute AFP with $\tau_r(\text{AdjFP} + \text{Language}) = +0.346$ and narrower CI.

The abnormal project identification process revealed that the two projects (P5 and P12)[1] are considered significantly abnormal. The results obtained confirm that attributes AdjFP and language are still the most influential attributes, with an improved correlation coefficient $\tau_r(\text{AdjFP} + \text{Language}) = +0.352$. No further abnormal project cases exist in the dataset. The attribute selected using EBA+ in here is different to the attributes selected by ANGEL that identified two attributes: Language and KSLOC, and Analogy-X that identified only AdjFP as predictive reliable attribute. However it is reasonable to believe that the combination of attributes

Added, AdjFP and Language will be useful for EBA. We can also notice that the categorical attribute Language is common between EBA+ and ANGEL, while attribute AdjFP is common between EBA+ and Analogy-X. This means that the attributes identified by EBA+ are also selected by ANGEL and Analogy-X which is an indication of their importance to effort prediction.

Table 8 Prediction accuracy comparison for Kemerer dataset

| Algorithm | Best Attribute set | MMRE | PRED(25)% | MdMRE |
|---|---|---|---|---|
| ALL | ALL | 73.7 | 26.7 | 55.2 |
| EBA+ | AdjFP + Language (with two projects removed (P5 and P12)) | **56.6** | **44.6** | **33.3** |
| ANGEL | Language + KSLOC | 63.8 | 40.0 | 33.3 |
| Analogy-X | AdjFP | 68.1 | 20 | 55.2 |

From Table 8, likewise Albrecht dataset, the best prediction accuracy was obtained by EBA+ in terms of MMRE and PRED(25) and MdMRE. As mentioned earlier, the target is to select the attributes that have sound statistical significance rather than optimizing based on some performance indicators. To assert that, Wilcoxon statistical sum rank test confirmed that the difference is not statistically significant as shown in Table 9 and Figure 9. From the results obtained, differences in absolute residuals of predictions between EBA+ and ANGEL are statistically insignificant, which means that median of both samples is not significantly different and their estimates are approximately similar.

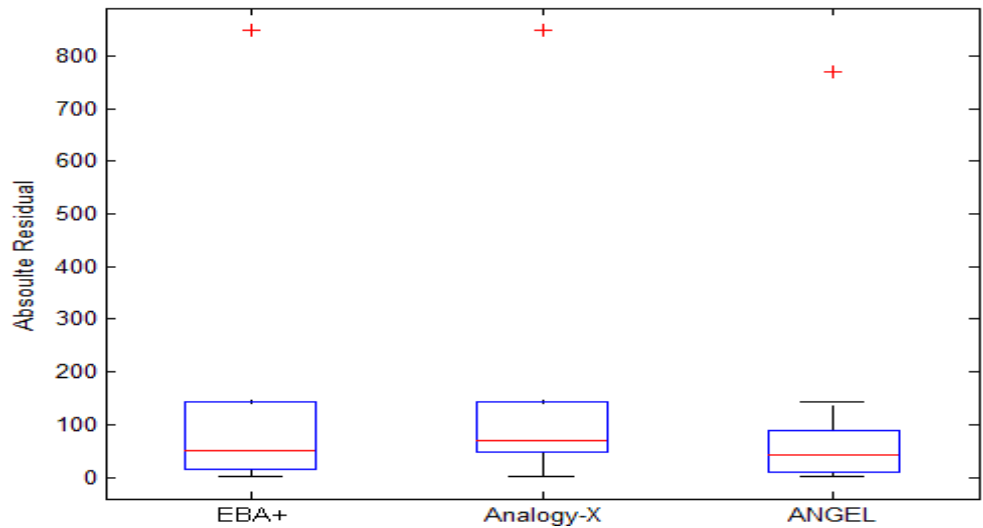

**Figure 9** Box-plot of absolute residuals of prediction using EBA+, Analogy-X, and ANGEL for Kemerer dataset

Table 9 Wilcoxon sum rank test for Kemerer dataset

| Algorithms | p-value | Rank sum |
|---|---|---|
| EBA+ Vs. ANGEL | 0.454 | 251 |
| EBA+ Vs. Analogy-X | 0.983 | 231.5 |

### 6.3.2 Desharnais dataset

The Desharnais dataset originally consists of 81 software projects collected from Canadian software houses [7]. This dataset is described by 10 attributes, two dependent attributes which are duration and the effort measured in 'person-hours', and 8 independent attributes. Unfortunately, 4 projects out of 81 contain missing values therefore we excluded them because they may be misleading the estimation process. This data pre-processing stage resulted in 77 complete software projects.

The first stage of EBA$^+$ method identified five significant influential attributes, namely: Transactions, Entities, RawFP, AdjFactor, AdjFPs and Dev.Env, for which their similarity matrices are statistically significant and correlated with effort similarity matrix. Although some attributes are not strongly correlated, they are significant as confirmed by permutation test. For example, similarity matrix based Dev.Env attribute is not strongly correlated with similarity matrix based effort with $\tau_r(\text{Dev.Env}) = +0.086$, but the significance test indicates that similarity matrix is significant and should be considered. So if the fitness criterion is the correlation coefficient only, then other attributes can be considered as influential attribute regardless of their significance. However, to ensure which attributes are more reliable for EBA, the significant attributes are then transferred into the next step in order to identify any possible attribute combination that are strong correlated than best individual attribute which is AdjFP with $\tau_r(\text{AdjFP}) = +0.278$ and CI=[0.2287  0.3173]. The next step identified AdjFP, and Dev.Env are the most influential attributes with $\tau_r(\text{AdjFP}+\text{Dev.Env}) = 0.312$ which is larger than $\tau_r(\text{AdjFP})$. The abnormal project identification process revealed that the three projects (P38, P70 and P76)[1] are considered significantly abnormal. The results obtained confirm that attributes AFP and Dev.Env are still the most influential attributes, with a slightly improved correlation coefficient $\tau_r(\text{AdjFP}+\text{Dev.Env}) = 0.332$, and no further abnormal project cases exist in the reduced dataset. To conclude the reduced model with selected attributes and projects are reliable to generate estimates using EBA.

Table 10 Prediction accuracy comparison for Desharnais dataset

| Algorithm | Best Attribute set | MMRE% | PRED(25)% | MdMRE% |
|---|---|---|---|---|
| ALL | ALL | 60.1 | 31.2 | 41.7 |
| EBA$^+$ | AdjFP+Dev.Env (with removed projects P76, P70 and P38) | **36.0** | **46.75** | 30.5 |
| ANGEL | AdjFP + Dev.Env | 38.2 | 42.9 | 30.8 |
| Analogy-X | AdjFP+Dev.Env (with a removed project P77) | 37.9 | 43.4 | **30.4** |

Unsurprisingly, ANGEL and Analogy-X also identified the same two influential attributes but with one abnormal project being removed by Analogy-X. Validation of analogy prediction using EBA$^+$ best attributes, ANGEL and Analogy-X are given in Table 10. Although all methods identified exactly the same attributes but EBA$^+$ beats ANGEL and Analogy-X in terms of MMRE because three abnormal projects were removed from the dataset. These results are also confirmed by Wilcoxon sum rank test on absolute residuals of predictions and Boxplots of residuals as shown in Table 11 and Figure 10 respectively. The results show also that the predictions generated by either of those attribute combinations will produce very similar estimates with p-value > 0.05, which indicates that the differences are not statistically significant. Furthermore, the median of absolute residuals of EBA$^+$ in Figure 10 is much smaller than other which confirms better predictions produced by EBA$^+$.

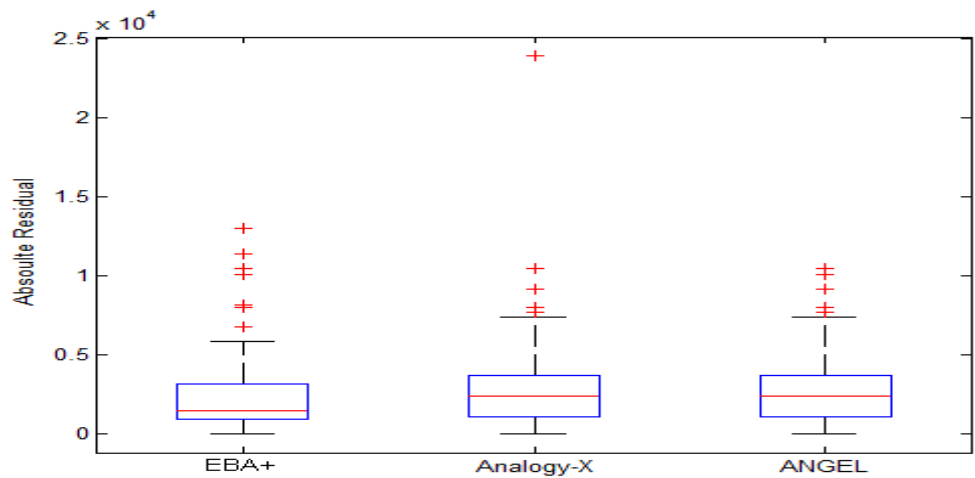

**Figure 10** Box-plot of absolute residuals of prediction using EBA$^+$, Analogy-X, and ANGEL for Desharnais

**Table 11** Wilcoxon sum rank test for Desharnais

| Algorithms | p-value | Rank sum |
|---|---|---|
| EBA$^+$ Vs. ANGEL | 0.287 | 5210 |
| EBA$^+$ Vs. Analogy-X | 0.264 | 5170 |

### 6.3.3 COCOMO dataset

The dataset COCOMO [7] was frequently used for validating various effort estimation methods. It includes 60 software projects that are described by 17 attributes in conjunction with an actual effort. The actual effort in the COCOMO dataset is measured by person-month which represents the number of months that one person needs to develop a given project. Despite the fact that the COCOMO dataset are now over 25 years old, it is still commonly used to assess the accuracy of new techniques.

Following a similar procedure with the previous cases above on COCOMO dataset, EBA$^+$ identified 5 significant correlated attributes from stage 1 are: TOOL, DATA, TURN, CPLX and LOC. The top significant attribute was LOC with $\tau_r(LOC)=0.6925$ with CI=[0.6640  0.7147]. Applying stage 2 of EBA$^+$ attribute selection on the significant correlated attributes resulted in identifying the similarity matrix based combination of LOC, TOOL and TURN as the most influential attributes with improved correlation coefficient $\tau_r(LOC+TOOL+TURN) = 0.72$ which is larger than $\tau_r(LOC)$ and with narrower CI.

The abnormal project identification process revealed that the three projects (P14, P54 and P57)[1] are considered significantly abnormal. The results obtained confirm that attributes LOC, TOOL and TURN are still the most influential attributes with no further abnormal project cases exist in the dataset.

Validation of analogy predictions using ANGEL obtained attributes (TIME, VEXP and LOC), EBA$^+$ reliable attributes (TURN, TOOL and LOC) and Analogy-X best attribute (LOC) are presented in Table 12. The results of EBA$^+$ are better than ANGEL and Analogy-x in terms of MMRE and PRED(25). This is confirmed by Wilcoxon sum rank significance test which shows that the differences between absolute residuals of predictions are statistically insignificant as shown in Table 13. Figure 11 Shows that EBA$^+$ generates good estimates with absolute residuals skewed to zero.

**Table 12** Prediction accuracy comparison for COCOMO dataset

| Algorithm | Best Attribute set | MMRE% | PRED(25)% | MdMRE% |
|---|---|---|---|---|
| ALL | ALL | 64.3 | 38.33 | 38.1 |
| EBA$^+$ | TURN+TOOL+LOC (with 3 removed projects (P14, P54 and P57)) | **26.9** | **56.7** | **21.5** |
| ANGEL | TIME+VEXP+LOC | 29.4 | 50.0 | 25.8 |
| Analogy-X | LOC | 44.9 | 36.7 | 34.9 |

**Table 13** Wilcoxon sum rank test for COCOMO

| Algorithms | p-value | Rank sum |
|---|---|---|
| EBA$^+$ Vs. ANGEL | 0.64 | 3542 |
| EBA$^+$ Vs. Analogy-X | 0.39 | 3247 |

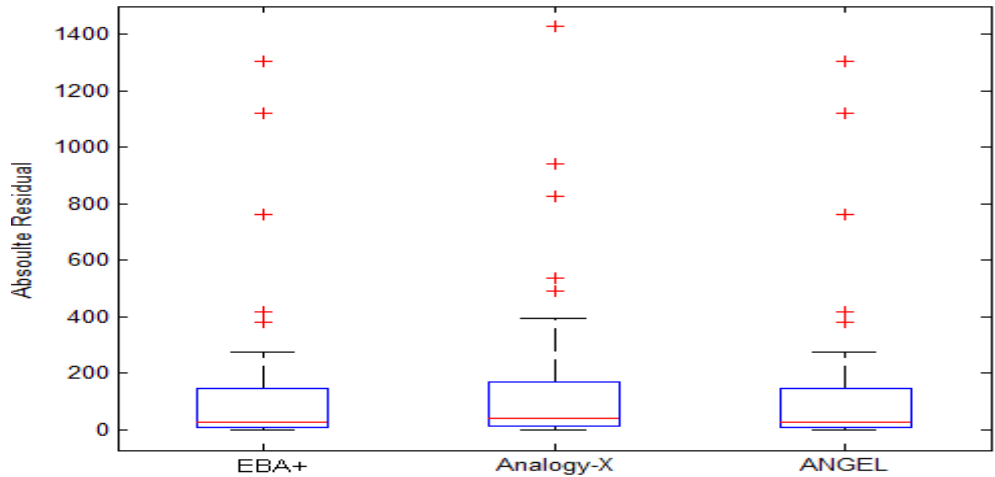

**Figure 11** Box-plot of absolute residuals of prediction using EBA$^+$, Analogy-X, and ANGEL for COCOMO

### 6.3.4 ISBSG dataset

The ISBSG Repository (Release 10) **[12]** currently contains more than 4,000 software projects gathered from different worldwide software development companies. All projects involved in the ISBSG repository are described by several numerical and categorical attributes. In order to assess the efficiency of the proposed similarity measures on software cost estimation we have selected a subset of attributes. Since many projects have missing values, only 575 projects with quality rating "A" are considered; 9 useful attributes were selected, 8 of which are numerical attributes and one is a categorical attribute. The selected attributes are: Adjusted Function Points (AFP), Input Functions (INC), Output Functions (OUC), Enquiry Functions (EQC), Count of Files (FILE), Interface Functions (INF), Added Functions (ADD), Changed Functions (CHC), and categorical attribute Resource Level (RSL).

In this section we investigate the effectiveness of EBA$^+$ on the large ISBSG dataset. As in other datasets we applied EBA$^+$ on ISBSG dataset, the first stage identified 6 significant correlated similarity matrices based on: AFP, OUC, INF, ADD, CHC and RSL individually. The best correlated attribute was $\tau_r(\text{OUC}) = 0.314$ with CI=[0.223  0.413]. It is interesting to note that the similarity matrix based on categorical attribute RSL (identified also by ANGEL), presents a stronger correlation than other continuous attributes such as INC and EQC. This indicates the capability of EBA$^+$ to identify the more influential attribute, either continuous or categorical.

The significant attributes are then transferred to the next stage in order to identify the best attribute combination, if exists. In this stage, it is found that the similarity matrix based attribute combination: AFP, OUC, INF and ADD produced a correlation value larger than $\tau_r(\text{OUC}) = 0.314$, with $\tau_r(\text{AFP} + \text{OUC} + \text{INF} + \text{ADD}) = +0.334$ and narrower CI. Therefore these attributes are regarded as the reliable predictive attributes. Furthermore, no abnormal projects have been identified. However, it is interesting to note that ANGEL identified 6 predictive attributes as shown in Table 14, while Analogy-X identified just two predictive attributes: "AFP" and "ADD".

In order to compare the results of different attribute selection algorithms, the best attribute subset obtained from each algorithm is then validated in terms of all performance indicators. For prediction accuracy comparison we used 10-fold cross-validation because the use of Jackknife is computationally far too intensive for a large dataset. The corresponding predictions accuracies in Table 14 show unsurprisingly that prediction accuracy of EBA$^+$ is comparable to ANGEL, and is better than prediction accuracy of Analogy-X. So we come to conclude that although the inclusion of relevant information affected the estimates, and consequently affected the level of prediction accuracy positively, but also the model acceptance is actually increased because the procedure of selection has a strong statistical background.

**Table 14** Prediction accuracy comparison for the ISBSG dataset

| Algorithm | Best Attribute set | MMRE% | PRED(25)% | MdMRE% |
|---|---|---|---|---|
| ALL | ALL | 74.4 | 35.8 | 43.8 |
| EBA$^+$ | AFP+OUC+INF+ADD | 65.7 | **44.2** | **32.3** |
| ANGEL | INC+OUC+FILE+INF+ADD+CHC | **59.2** | 43.4 | 34.2 |
| Analogy-X | AFP+ADD | 70.6 | 39.6 | 39.4 |

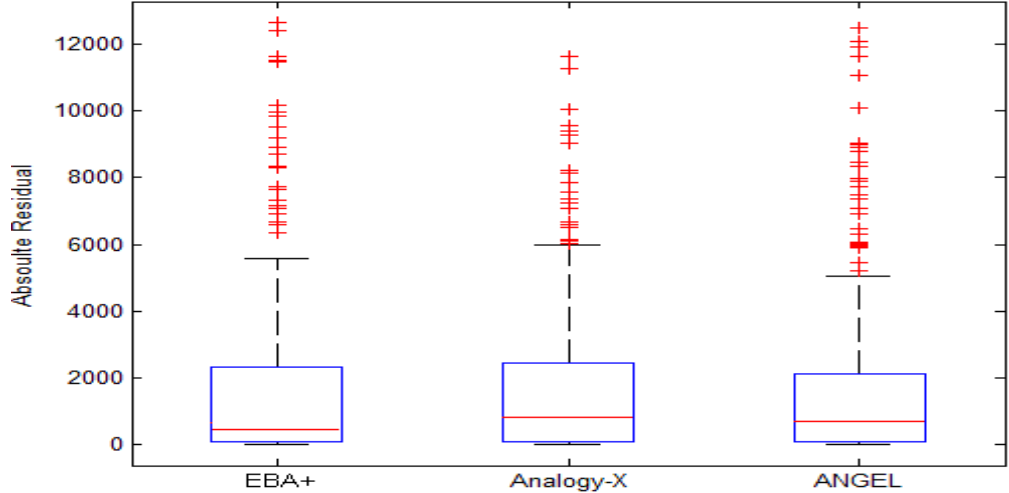

**Figure 12** Box-plot of absolute residuals of prediction using EBA$^+$, Analogy-X, and ANGEL for ISBSG

The significance test based on Wilcoxon sum rank test (shown in Table 15) suggests that the differences in predictions between EBA$^+$ and ANGEL, Analogy-X are insignificant. Further, Figure 12 shows that the absolute residuals of predictions generated by either of those attribute combinations will produce similar estimates, indicating that the differences are not statistically significant.

**Table 15** Wilcoxon sum rank test for ISBSG

| Algorithms | $p$-value | Rank sum |
|---|---|---|
| EBA$^+$ Vs. ANGEL | 0.321 | 249780 |
| EBA$^+$ Vs. Analogy-X | 0.853 | 251342 |

## 7 Discussion and Conclusions

Quality of dataset that satisfies main assumption of an estimation model is very important for reliable effort predictions. EBA has fundamental hypothesis that governs the estimation process which assumes: The projects that are similar in terms of projects descriptors are also similar in terms of project effort values. This hypothesis implies that any attribute is regarded as reliable for prediction only if it has a strong relationship with the effort, otherwise it is irrelevant. Also it implies that for any target project the ranks of similar source projects with respect to a particular attribute should be consistent with ranks of similar source projects with respect to their effort values. Based on above assumptions we developed a new method to test quality of dataset using Kendall row wise correlation for EBA. Much of the initial work on EBA$^+$ was motivated by experiences with software effort estimation using analogy and based upon the ANGEL toolset developed in **[24]** and Analogy-X developed by Keung et al.**[16]**. ANGEL's attribute selection techniques have the limitation of being tightly coupled with the actual results, but provide no statistical evidence that the attributes satisfy fundamental hypothesis of EBA. On the other hand, Analogy-X cannot assess the relevancy of nominal attributes individually; however this can be done after identifying best numerical attributes and then partitioning dataset into subgroups according to the number of categories by only considering matrix elements of cases within the same group. However this approach for handling nominal attributes is not efficient, and removes a large number of matrix elements in the similarity matrix. Therefore EBA$^+$ is designed to overcome the main limitations of ANGEL and Analogy-X.

The proposed method consists mainly of two processes, the first one attempts to select all reliable attributes that significantly satisfy EBA hypothesis, whilst the second process attempts to identify abnormal projects that undermines estimation process. If the test procedure of EBA$^+$ failed to identify any reliable attributes (i.e. null hypothesis cannot be rejected) then the dataset is no longer reliable to generate predictions using EBA. However, using Kendall's row-wise rank correlation gives a clear picture of how ranks of closest projects to the target project are considered important in judging which attribute is more reliable and influential than others. The results derived from EBA$^+$ can be used to assess whether analogy is indeed appropriate for the given dataset or not. EBA$^+$ method can be easily incorporated into any analogy estimation tools such as ANGEL **[24]**, as an extension or a plug-in. The fundamental procedures can be fully automated in the tools, and would not be visible to the user other than to advise the user when analogy was not appropriate for the dataset under investigation. In summary, EBA$^+$ functionality are:
- It assesses the appropriateness of a specific dataset for EBA.
- It provides a statistical mechanism for attribute subset selection.
- It is able to identify the appropriateness of categorical attributes for EBA.
- It provides a statistical mechanism to identify abnormal projects within a dataset.

It has been shown that EBA$^+$ greatly improves the performance of EBA. EBA$^+$ works at a preliminary stage in an analogy-based system, the dataset preparation and evaluation level. There is no doubt that an evaluation system at this level significantly influences the later process and thus the prediction outcome. Performance studies on six established datasets indicated that EBA$^+$ is highly competitive when compared with ANGEL's brute-force search and Analogy-X algorithms in term of prediction accuracy and statistical significance. Furthermore, EBA$^+$ scales well if compared with ANGEL's brute-force searching algorithm with regards to computation power. However, results showed that EBA$^+$ may take longer to compute than expected especially when dealing with large datasets such as ISBSG because of the Bootstrapping procedure which is computationally far too intensive.

Further research is necessary to assess the overall efficiency and performance of EBA$^+$ on datasets with many categorical attributes. In the future work, we will extend EBA$^+$ to include attribute weighting derivation. We will also study the problem of using categorical data as nominal and ordinal scale data type and their impact on attribute weighting.

## 8 Acknowledgements

The authors are grateful to the Applied Science Private University, Amman, Jordan, for the financial support granted to this research project.